\documentstyle[aps,preprint,epsfig,amssymb,floats]{revtex}
\begin{document}
\draft

\preprint{UAB--FT--416}

\title{\Large {\bf Constraints on the mass of the superlight gravitino\\from the muon anomaly}}

\vspace{2in}
\author{F.Ferrer and J.A.Grifols}

\vspace{.5in}

\address{Grup de F{\'\i}sica Te\`orica and Institut de F{\'\i}sica d'Altes Energies,\\
Universitat Aut\`onoma de Barcelona, E--08193 Bellaterra, Barcelona, Spain}
\date{April 1997}
\maketitle

\vspace{1in}

\begin{abstract}
We reexamine the limits on the gravitino mass supplied by the muon anomaly in the frame of supergravity models with a superlight gravitino and a superlight scalar $S$ and a superlight pseudoscalar $P$.
\end{abstract}

\vspace{1in}
\pacs{PACS number(s): 14.80.Ly, 13.40.Em, 04.65.+e, 12.60.Jv}

In a wide class of supergravity models with SUSY breaking scale in the TeV range, the gravitino can be very light $(m_{3/2} \sim M^2_{SUSY}/M_{Pl})$. In fact, its mass could lie anywhere between $\mu eV$ and $keV$. Examples for this are those models where gauge interactions mediate the breakdown of supersymmetry \cite{dine} or models where an anomalous $U(1)$ gauge symmetry induces SUSY breaking \cite{bine}. Also, no--scale models can accomodate superlight gravitinos \cite{ellis}.

Clearly, it is important to bound and eventually determine the mass of the gravitino. To mention only an instance where the gravitino mass is of great physical significance, let us recall that a mass on the order of a few $keV$ can be relevant for the dark matter problem. The sources of direct (laboratory) information on the gravitino mass are rare \cite{bhat,roy,mendez,aguila} and perhaps the best one comes from the $(g-2)_{\mu}$ of the muon \cite{mendez,aguila}. Of course, more indirect information is available from cosmological and astrophysical settings \cite{fukugita}. Here, we shall be exclusively concerned with the bounds obtained from the muon anomaly.

The first calculation of the gravitino contribution to $(g-2)_\mu$ gave $m_{3/2}\gtrsim 10^{-6}\:eV$ for a smuon mass of ${\cal O}(100\: GeV)$. However,this bound was obtained in the context of a supergravity model with a massless photino and no other light particles other than the gravitino present. The case with a massive photino was considered later by {\it del Aguila} \cite{aguila}. In the general nonzero photino mass, the contribution to $(g-2)_{\mu}$ in broken supergravity is not finite. Nevertheless, it is finite in models with a constant K\"{a}hler metric, which is the case for the models under consideration in our present work. Again, the bounds obtained in this case for the gravitino mass were on the order of the previously found.

But, it is also a generic feature of some of the recently considered models that the superlight gravitino is accompanied by a superlight scalar $S$ and pseudo--scalar $P$ particles. These particles and the gravitino are coupled to matter with strength inversely proportional to the gravitino mass and, hence, their effects are magnified for sufficiently low gravitino masses. In this brief report we reexamine the issue of the phenomenological bounds on the mass of the gravitino in this general class of models. In particular, we include in our estimates the effect on the bound caused by the $S$ and $P$ particles, a contribution that had not been considered before.

The amplitudes involving gravitino exchange that contribute to leading order in Newton's constant $G_N$ have been already computed in \cite{mendez,aguila}, as we commented above, and we just use the results. Therefore, we need only to compute the extra diagrams that involve $S$ and $P$ particles to leading order in the coupling strength. The effective lagrangian governing the interactions of matter and gauge fields with gravitinos and scalars $S$ and $P$ can be found in the literature. The explicit form of the $S$ and $P$ couplings reads \cite{cremmer}:
\begin{equation}
e^{-1} {\cal L}=\sqrt{\frac{\pi G}{3}}\left(\frac{m_{\tilde{\gamma}}}{m_{3/2}}\right) \left( S F^{\mu\nu}F_{\mu\nu}+P\tilde{F}^{\mu\nu}F_{\mu\nu}\right)
\end{equation}
The $S/P$ couplings to leptons do not contain the $m^{-1}_{3/2}$ factor and are therefore neglegible. This is why their contribution to $(g-2)_{\mu}$ was neglected in \cite{aguila}. But we should consider the diagrams shown in Fig. 1 that include the virtual exchange of $S$ and $P$ particles and are many orders of magnitude larger than the diagrams where $S/P$ are attached to the muon line. 

\begin{figure}[bht]
\begin{center}
\epsfig{file=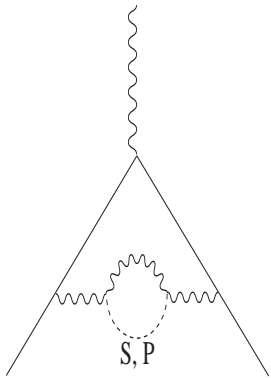,width=3.5cm,height=5cm}
\end{center}
\caption{\it $S$ and $P$ contributions to $a_{\mu}$}
\end{figure}

From inspection of the diagram we see that this contribution is proportional to $\alpha\,G_N\left(\frac{m_{\tilde{\gamma}}}{m_{3/2}}\right)^2$, with $\alpha$ the fine--structure constant and might numerically compete with the leading gravitino contribution.  We have calculated it following the method outlined in \cite{rafael}:
\begin{equation}
a_{\mu}\equiv\frac{(g-2)_{\mu}}{2}=\frac{1}{\pi}\int_0^\infty{\frac{dt}{t}I\!m\, \Pi^{S,P}(t)\,K(t)}
\end{equation}
where the function $K(t)$ is given by
\begin{equation}
K(t)=\frac{\alpha}{\pi}\int_0^1{dz \frac{z^2 (1-z)}{z^2+t(1-z)/\mu^2}}
\end{equation}
where $\mu$ is the mass of the muon. Its explicit form can be found in \cite{rafael}.

The function $\Pi^{S,P}(t)$ appearing in (2) is the vacuum polarization insertion shown in Fig. 1. The explicit form for it can be obtained using (1). We find,
\begin{equation}
I\!m\, \Pi^{S,P}(t)=\frac{G_N}{18}\left(\frac{m_{\tilde{\gamma}}}{m_{3/2}}\right)^2 t (1-m^2_{S,P}/t)^3
\end{equation}
This form for $I\!m\, \Pi^{S,P}$, when introduced in (2), renders the integral U.V. infinite. This is a result of unitarity violation by amplitude (4) which traces back to the nonrenormalizability of the supergravity lagrangian. Indeed, above a critical energy,
\begin{equation}
E\sim {\cal O} \left( m_{3/2}/G^{1/2}_N m_{\tilde{\gamma}}\right)
\end{equation}
the amplitude (4) no longer respects the unitarity limit.

For $m_{\tilde{\gamma}}\sim {\cal O}(100\,GeV)$ and a very light gravitino $(\mu eV)$, E is ${\cal O}$(few TeV). Unitarity demands (4) to be bound by a constant. So, we may choose to replace (4), for $t\geq E^2$, by the constant value that saturates the unitarity limit. A phenomenologically equivalent procedure, is to cut--off the integral (2) at the critical energy (5). Moreover, since the divergence is only logarithmic, it really makes no appreciable difference which precise value of the cut--off we use.

As long as $m_{S,P} \ll \mu$, the effect of the mass of the scalars is negligible. Hence, in our numerical analysis we set $m_{S,P} = 0$.
 
The allowed discrepancy between experimental value and predicted theoretical (SM) anomaly for the muon is at $90\%\:C.L.$ \cite{carena}:
\begin{equation}
-9\cdot 10^{-9}\leq \Delta a_{\mu}\leq 19\cdot 10^{-9}.
\end{equation}
We can saturate $\Delta a_{\mu}$ given by (6) with the supergravity contributions discussed here to obtain the advertised constraints on $m_{3/2}$.

As mentioned before, we borrow from \cite{aguila} the contribution of the gravitino diagrams:
\begin{eqnarray}
a_{\mu}^{3/2}=\frac{G_N}{6 \pi}\mu^2 \sum_{k=1,2}^2\left[\frac{1}{6}+\frac{m_{\tilde{\gamma}}^2}{\tilde{m}_k^2-m_{\tilde{\gamma}}^2} \left(\frac{m_{\tilde{\gamma}}^2}{\tilde{m}_k^2-m_{\tilde{\gamma}}^2} \log \frac{\tilde{m}_k^2}{m_{\tilde{\gamma}}^2}-1\right) + (-1)^k \sin 2\, \theta \frac{m_{\tilde{\gamma}}}{\mu} \right]\left(\frac{\tilde{m}_k}{m_{3/2}}\right)^2 
\end{eqnarray}
here, $\tilde{m}_1$, $\tilde{m}_2$, and $\theta$ are the masses and mixing angle, respectively, of the scalar partners of the muon.

To (7) we must add the contribution from Fig. 1, as obtained from (2), (3) and (4). The total (gravitino $\mbox{}+S+P$) supergravity contribution to $(g-2)_{\mu}$ depends then on $m_{3/2}$, $m_{\tilde{\gamma}}$, $\tilde{m}_1$, $\tilde{m}_2$, and $\theta$. The bound on $m_{3/2}$ that follows from the experimental requirement on (6) thus depends on 4 parameters. A general analysis in parameter space is outside the scope of this paper. Rather, we wish to assess a conservative order of magnitude estimate for the bound on the gravitino mass, in the general framework of supergravity models endowed with superlight particles, that follows from the muon anomaly. In this spirit, we consider maximal mixing, i.e. $\theta=\pi/4$ and $m_{\tilde{\gamma}}=100$  GeV, which is a typical figure in these kind of models. To present plots, we trade the parameters $\tilde{\mu}\equiv \frac{\tilde{m}_1+\tilde{m}_2}{2}$ and the splitting $\Delta \equiv \frac{\tilde{m}_1-\tilde{m}_2}{\tilde{\mu}}$ for $\tilde{m}_1$ and $\tilde{m}_2$. Fig. 2 displays the allowed gravitino mass values as a function of $x=\frac{\tilde{\mu}}{m_{\tilde{\gamma}}}$ and for $\Delta=0\%, 25\%, 100\%, 300\%$.

We see from these plots that the gravitino mass should be above $10^{-6}$ to $10^{-4}\:eV$, for the phenomenologically interesting smuon mass interval $50$ to $500\:GeV$, to safely comply with the constraints posed by the muon anomaly.

Let us end with the comment that the E821 experiment at BNL with an expected improvement in precision by one to two orders of magnitude will correspondingly improve the limits on the gravitino mass derived in this paper.

\begin{center}
{\large {\bf Acknowledgements}}
\end{center}
\vspace{-0.2in}
Work supported by the CICYT Research Project AEN95--0882. F.F. acknowledges the CIRIT for financial support.

\begin{figure}[hhh]
\begin{center}
\begin{tabular}{cc}
\epsfig{file=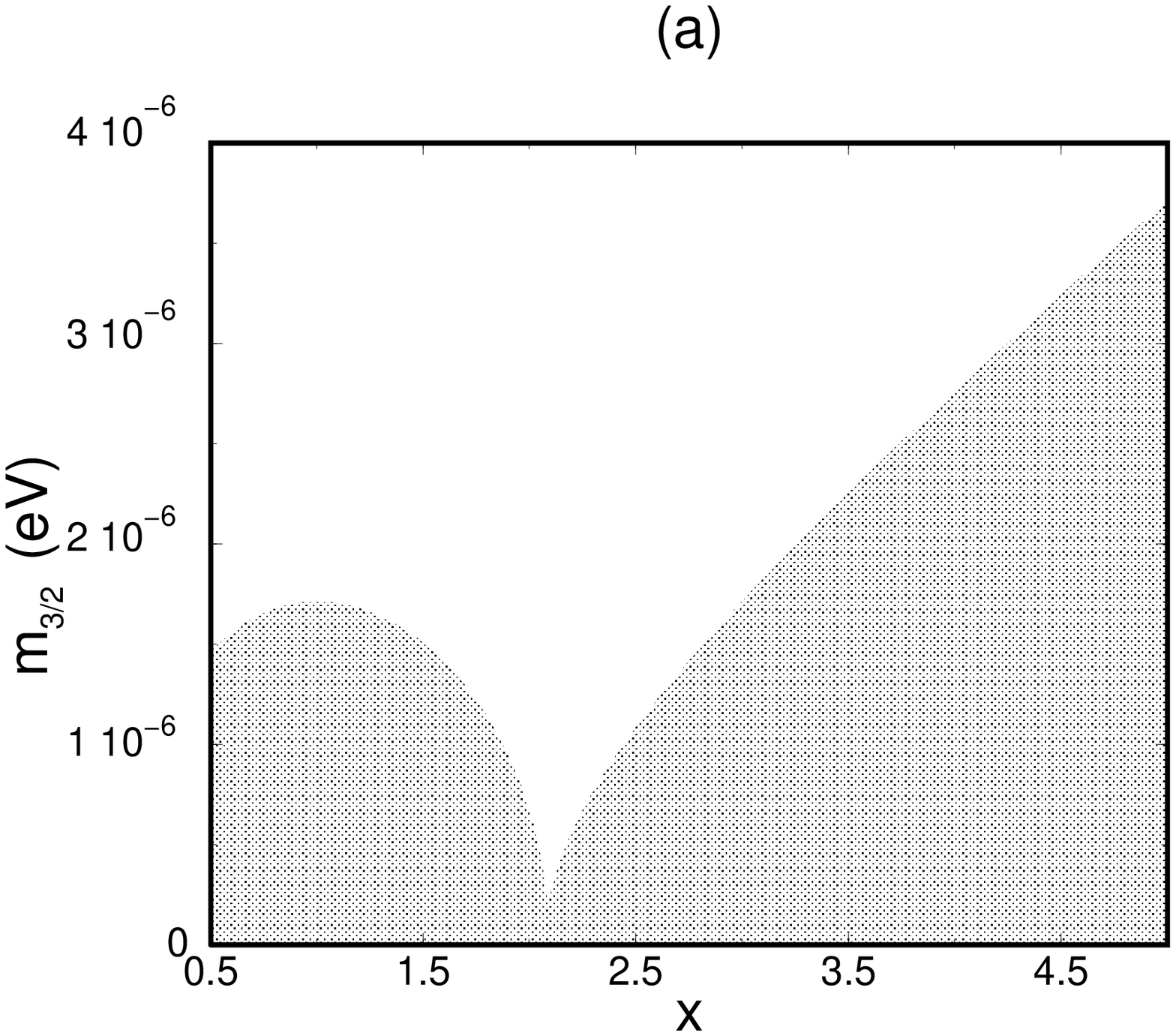,width=6cm,height=6cm} & 
\epsfig{file=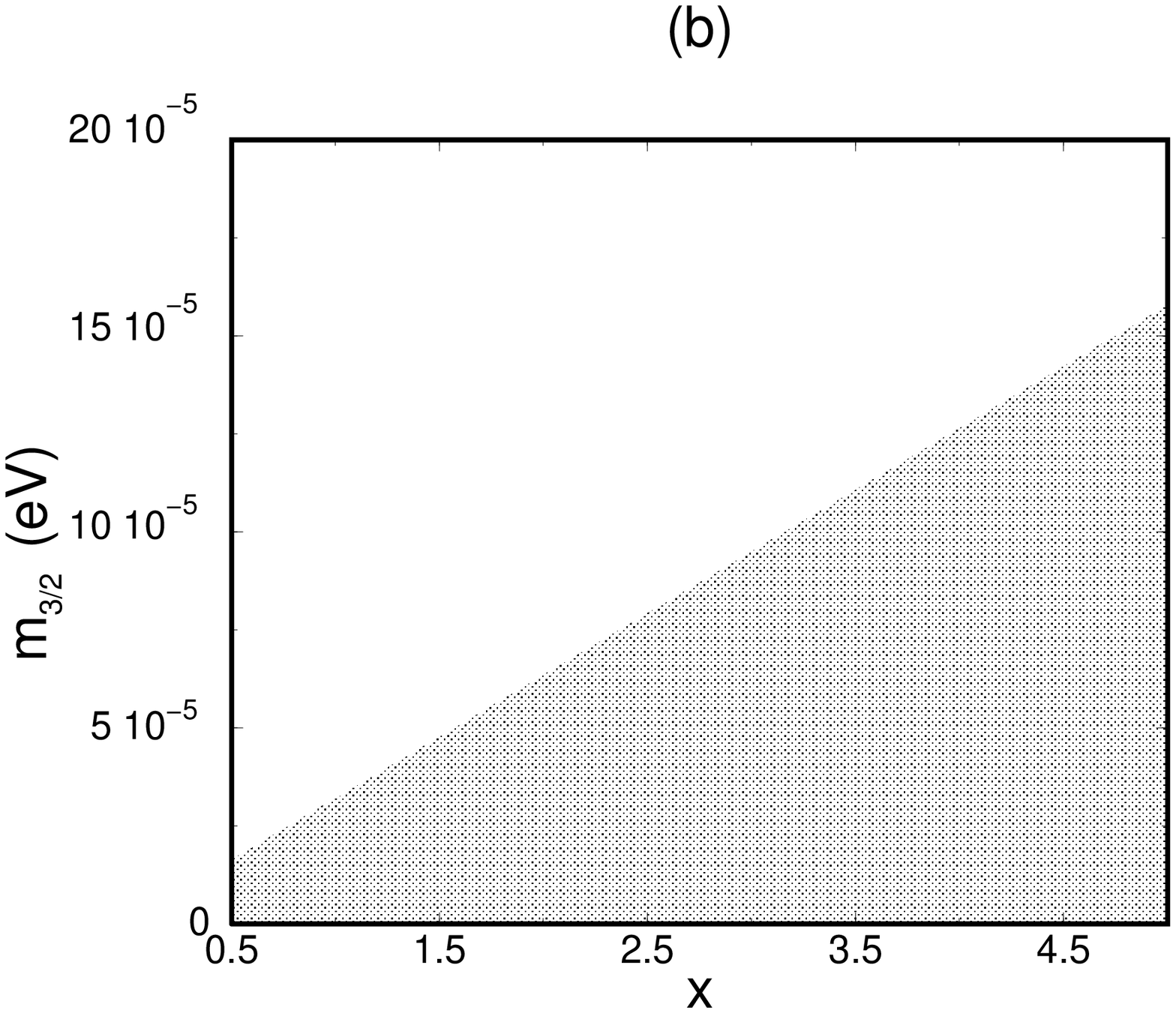,width=6cm,height=6cm} \\
\epsfig{file=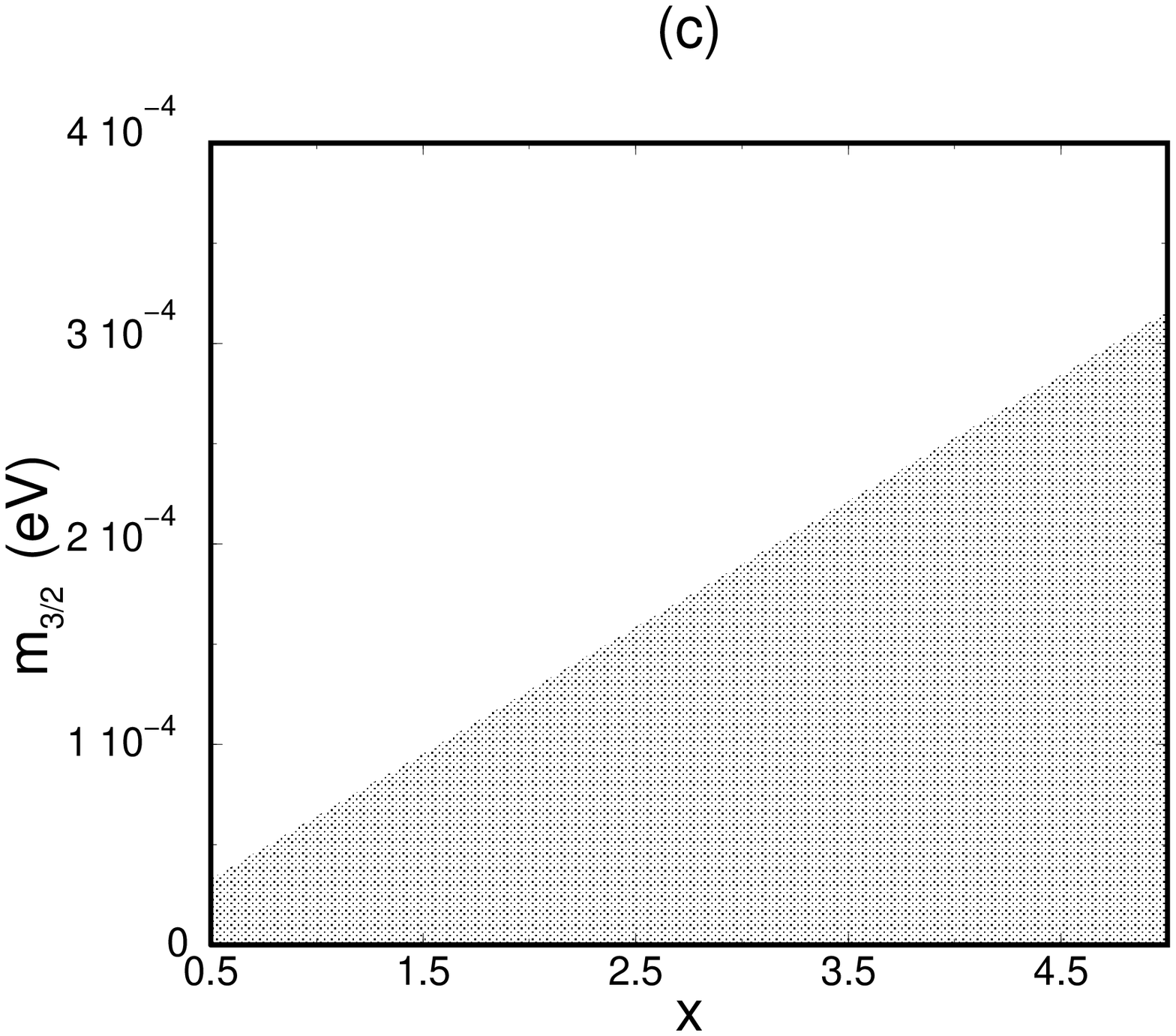,width=6cm,height=6cm} & 
\epsfig{file=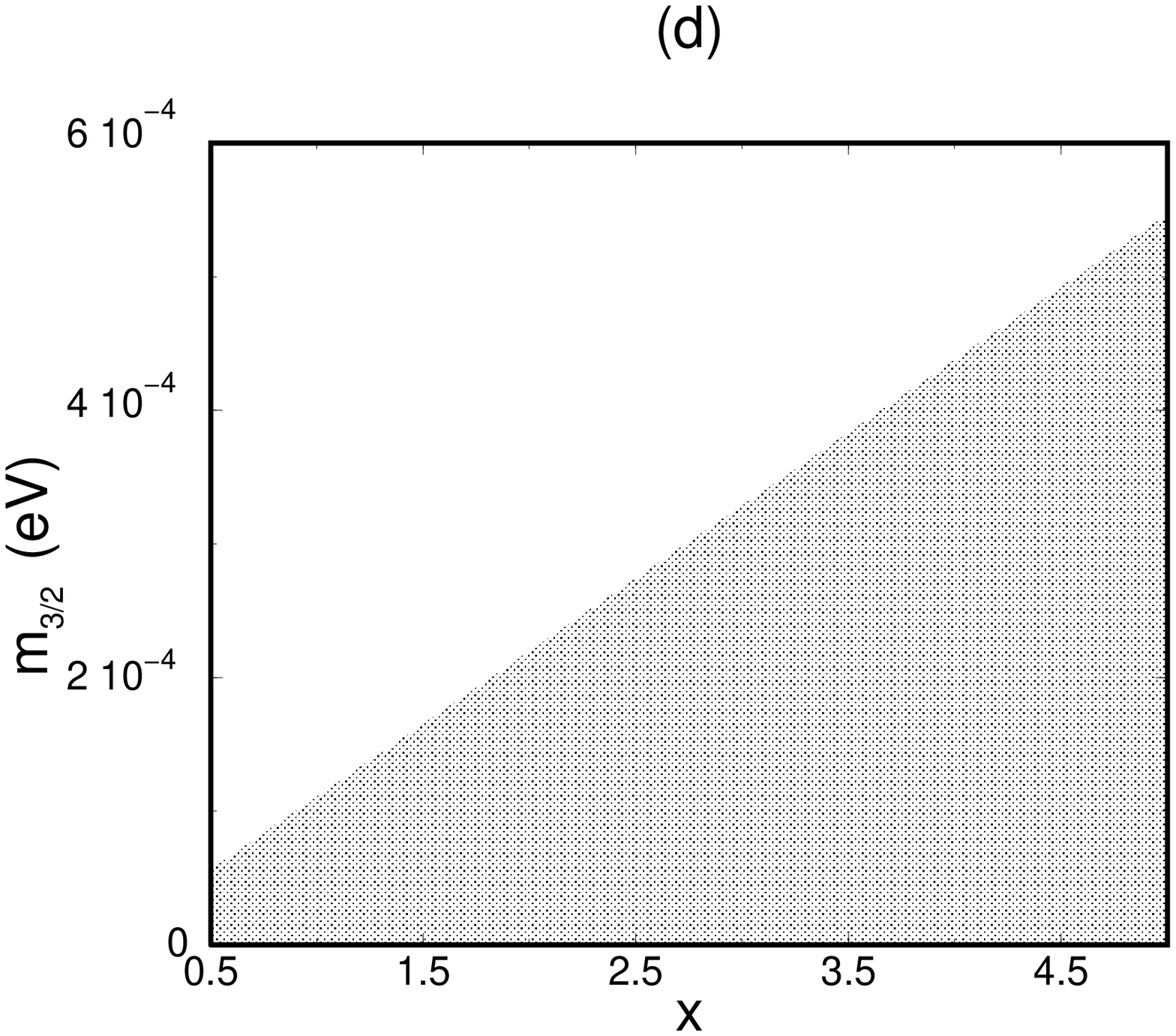,width=6cm,height=6cm} \\
\end{tabular}
\end{center}
\caption{\it Bound on $m_{3/2}$. Shaded area is excluded by $a_{\mu}$. Shown are the results for four different splittings $\Delta$: (a) 0\%, (b) 25\%, (c) 100\%, (d) 300\%}
\end{figure}

\end{document}